\documentclass[preprint,12pt]{elsarticle}

\usepackage{amssymb}

\usepackage{hyperref}
\biboptions{sort&compress}

\journal{Journal of Magnetism and Magnetic Materials}

\begin{document}

\begin{frontmatter}



\title{Generalized analysis of thermally activated domain-wall motion in Co/Pt multilayers}


\author[mit,neu]{Satoru Emori}
\ead{s.emori@neu.edu}
\author[mit]{Chinedum K. Umachi}
\author[mit]{David C. Bono}
\author[mit]{Geoffrey S. D. Beach}

\address[mit]{Department of Materials Science and Engineering, Massachusetts Institute of Technology, Cambridge, MA 02139, US}
\address[neu]{Department of Electrical and Computer Engineering, Northeastern University, Boston, MA 02115, USA}
\begin{abstract}
Thermally activated domain-wall (DW) motion driven by magnetic field and electric current is investigated experimentally in out-of-plane magnetized Pt(Co/Pt)$_3$ multilayers.  We directly extract the thermal activation energy barrier for DW motion and observe the dynamic regimes of creep, depinning, and viscous flow.  Further analysis reveals that the activation energy must be corrected with a factor dependent on the Curie temperature, and we derive a generalized Arrhenius-like equation governing thermally activated motion.  By using this generalized equation, we quantify the efficiency of current-induced spin torque in assisting DW motion.  Current produces no effect aside from Joule heating in the multilayer with 7-\AA\ thick Co layers, whereas it generates a finite spin torque on DWs in the multilayer with atomically thin 3-\AA\ Co layers.  These findings suggest that conventional spin-transfer torques from in-plane spin-polarized current do not drive DWs in ultrathin Co/Pt multilayers.  

\end{abstract}


\end{frontmatter}


\section{\label{intro}Introduction}

The dynamics of magnetic domain walls (DWs) driven by magnetic field~\cite{ferre, lemerle, cayssol, kimNat, metaxas, lavrijPin, emoriMOKE, jeDMI, wuth, allwood, kiermaier} or electric current~\cite{parkin, currivan, koyamaNatMat, ueda, tanigawa, ngo, koyamaHindep, mooreHighSpeed, mironDW, ryuParkin, koyamaMgO, ravelosona, boulle, burrowes, alvarez, heinen, kimAPEX, leePRL, mironSTMeter, suzuki, cormierPRB, cormierJPD, lavrijPt, kimEa, kimDHtemp, kimTwoBarrier, emoriJPCM, yamanouchi, curiale, haazen, emoriGdOx, emoriChiral, ryuChiral, emoriDMI} is often dictated by interactions with defects.  For potential device applications~\cite{kiermaier, currivan, parkin, allwood}, it is important to investigate DW dynamics in patterned strips of ferromagnetic thin films, especially materials with perpendicular magnetic anisotropy (PMA) that have shown low critical current densities and high maximum velocities for DW motion~\cite{ueda, tanigawa, ngo, koyamaHindep, mooreHighSpeed, mironDW, ryuParkin, koyamaMgO, emoriGdOx, emoriChiral, ryuChiral, emoriDMI}.  DWs in these out-of-plane magnetized thin films are narrow ($\sim$1-10 nm wide) and susceptible to pinning by nanoscale defects at sufficiently low driving fields or current densities~\cite{ferre, lemerle, cayssol, kimNat, metaxas, lavrijPin, emoriMOKE, jeDMI, ravelosona, boulle, burrowes, alvarez, heinen, kimAPEX, leePRL, mironSTMeter, cormierJPD, lavrijPt, kimEa, kimDHtemp, kimTwoBarrier, emoriJPCM, emoriGdOx, yamanouchi, curiale, emoriChiral, kimTwoBarrier, haazen}.  In this case, a DW moves stochastically from one pinning site to another by thermal activation~\cite{ferre, lemerle, martinezStoch}.  Reliable device operation requires a good understanding of such stochastic DW dynamics.  

Thermally activated DW dynamics is also interesting from the standpoint of fundamental physics.  The dynamics of a DW driven through disorder can exhibit universal scaling behavior spanning several decades in velocity, with scaling exponents that depend on the sample dimensionality~\cite{kimNat} and the nature of the driving forces~\cite{leePRL, yamanouchi}.  Thermally activated DW motion is typically understood to follow the power law of creep, which relates the thermal activation energy barrier $E_A$ to the effective driving field $H_{eff}$ through a scaling exponent $\mu$.  In the case of a one-dimensional elastic DW in a two-dimensional disorder potential, appropriate for ultrathin films with PMA, it has been shown theoretically that $\mu = 1/4$~\cite{lemerle}. This form of creep scaling has been shown, or assumed, to hold in many other experimental studies~\cite{lemerle, cayssol, metaxas, kimNat, lavrijPin, mooreHighSpeed, alvarez, leePRL, cormierJPD, lavrijPt, kimDHtemp} of out-of-plane magnetized ultrathin films, typically evidenced by linearly fitting the logarithm of the DW velocity against  ${H_{eff}}^{-\mu}$.  However, a recent investigation~\cite{emoriJPCM} of thermally activated DW motion in Co/Pt multilayers with PMA indicates that such an analysis is relatively insensitive to $\mu$, and that significant deviations from universal scaling can become apparent when sample temperature is included as a variable.  To elucidate the fundamentals of thermally activated DW dynamics, a rigorous and generalized experimental scheme incorporating temperature dependence is needed.  

In general, the velocity of a DW in the thermally activated regime is expressed by the Arrhenius relationship, 
\begin{equation}
v = v_o \exp\left(\frac{-E_A}{k_{B}T}\right),
\label{eq1}
\end{equation}
where $k_{B}$ is the Boltzmann constant, $T$ is the sample temperature, and $v_o$ is the pre-exponential factor.  Because of this exponential relationship, the DW velocity (or depinning rate) is sensitive to even small variations in temperature.  Therefore, the DW velocity may increase significantly through Joule heating from driving current~\cite{suzuki, cormierPRB, cormierJPD, emoriJPCM}, obscuring the contributions from spin torque effects~\cite{ralph, beach, brataas, zhang, thiavilleSTT, tatara, khval}.  By contrast, the activation energy depends directly on the driving field $H$ and current density $J_e$, which can be considered together as an effective field (e.g.  $H_{eff} = H + \epsilon J_e + cJ_e^2...$)~\cite{leePRL, kimEa, kimJV, ryu}.  The efficiencies of spin torques ($\epsilon$, $c$, etc.) may be extracted by examining the functional dependence of the activation energy barrier on the driving current.  
This is often performed by analyzing the DW velocity within the framework of a particular creep scaling model.  However, $E_A$ can also be extracted directly from Eq. \ref{eq1} through an Arrhenius analysis~\cite{emoriJPCM}, from which its dependence on driving forces can be evaluated empirically without assuming a particular model of creep scaling. 
Despite the significance and utility of the activation energy, there have been few studies to measure it directly as a function of field and current through such a temperature-dependent analysis~\cite{emoriJPCM, emoriGdOx}.

We present a comprehensive study of field- and current-driven thermally activated DW motion in out-of-plane magnetized Co/Pt multilayers by building on the approach in~\cite{emoriJPCM}.  From temperature-dependent measurements spanning up to 8 decades in DW velocity, we directly extract the activation energy over a wide range of driving field, from deep in the creep regime up to the viscous flow regime.  Further analysis of the activation energy as a function of driving field reveals a nontrivial dependence on the Curie temperature of the sample.  By incorporating this newly found temperature contribution, we empirically derive a modified Arrhenius-like relationship that determines the DW velocity as a complete function of driving field and temperature.  From this Arrhenius-like relationship and the current dependence of the activation energy, we quantify the effects of current on DW motion as an effective driving field.  Our results demonstrate limitations of the universal creep scaling law and the robustness of the direct analysis of the activation energy. 

Furthermore, the spin-torque efficiencies for Co/Pt multilayers with different Co layer thicknesses resolve the disparity in recent studies of current-induced DW dynamics in Co/Pt, some reporting high spin-torque efficiencies of over 10 Oe/10$^{11}$ A/m$^2$~\cite{ravelosona, boulle, alvarez, leePRL, lavrijPt} while others showing no effects other than Joule heating~\cite{mooreHighSpeed, mironSTMeter, suzuki, cormierPRB, cormierJPD, lavrijPt, emoriJPCM, emoriGdOx}.  Our results suggest that the in-plane current through the ultrathin ferromagnetic layers attains a vanishingly small effective spin polarization, consistent with the rigorous semi-classical calculations by Cormier \emph{et al}.~\cite{cormierPRB} and the recent experimental findings by Tanigawa \emph{et al}.~\cite{tanigawa}  In ultrathin Co/Pt-based structures, conventional spin-transfer torques (STTs) are likely not present, and the current-induced torque from the spin Hall effect~\cite{liuPt, liuTa} drives DWs as reported in recent studies~\cite{haazen, emoriChiral, ryuChiral, emoriDMI, khval, thiavilleDMI}.

\section{\label{sec2}Experimental Details}
\begin{figure}
\centering
\includegraphics[width=0.5\columnwidth]{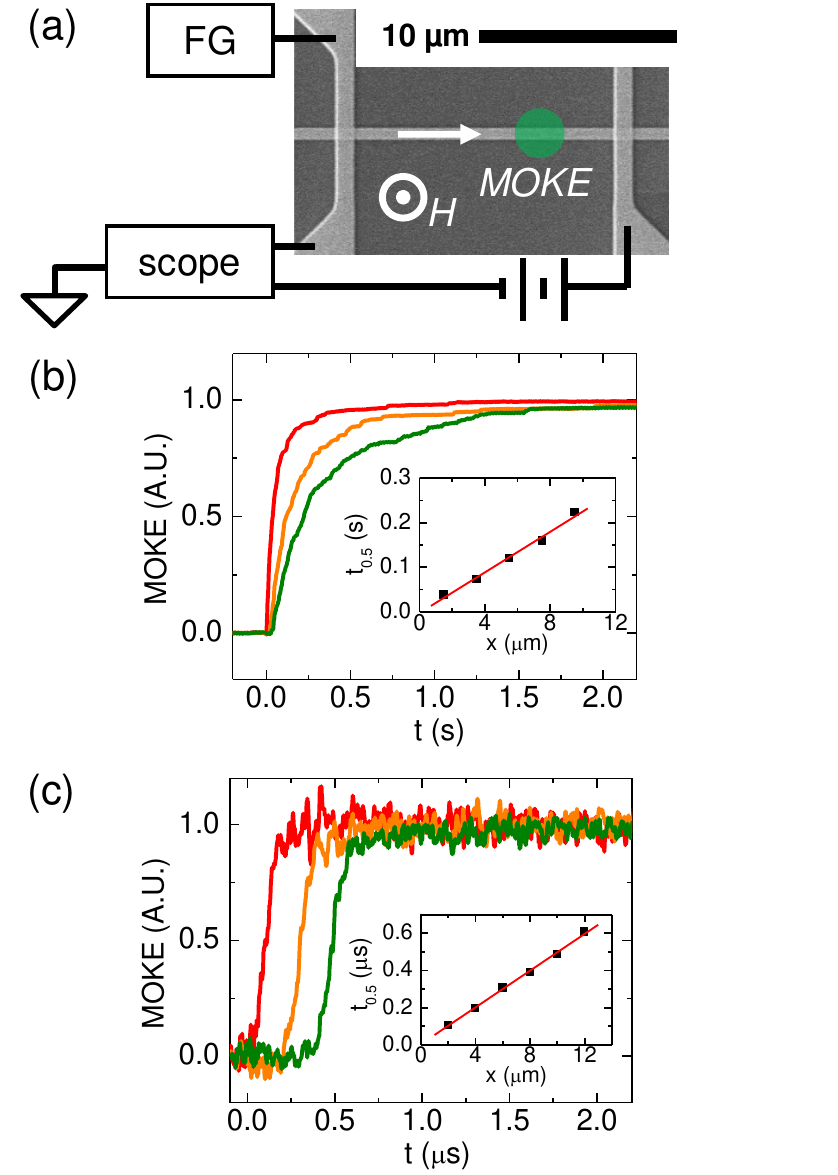}
\caption{(a) Scanning electron micrograph and measurement schematic of a Co/Pt sample.  (b, c) Examples of averaged MOKE transients for the $t_{Co}$= 7 \AA\ strip at $T$ = 300 K and (b) $H$ = 123 Oe and (c) $H$ = 230 Oe.   Individual curves are transients measured at different positions along the Co/Pt strip. The insets show linear fitting for average arrival time $t_{0.5}$ versus probed position.} 
\label{fig1}
\end{figure}

500-nm wide Co/Pt multilayer strips with electrodes (Fig. \ref{fig1}(a)) were fabricated using e-beam lithography, sputtering, and liftoff.  The multilayer structure was Si / SiO$_2$(500) / TaOx(40) / Pt(16) / [Co($t_{Co}$)/Pt(10)]$_{2}$/ Co($t_{Co}$)/ Pt(16), where numbers in parentheses indicate thicknesses in \AA.  The Pt layer thicknesses and TaOx underlayer were optimized in an attempt to maximize current flow through the ferromagnetic Co layers while maintaining strong PMA~\cite{emoriCoPt}.  We present results for $t_{Co}$ = 7 \AA\ and 3 \AA\, the upper and lower limits, respectively, at which the remanent magnetization was fully out of plane and the DW nucleation field $H_{nuc}$ exceeded the propagation field.  $H_{nuc}$ was $\approx$230 Oe for $t_{Co}$ = 7 \AA\ and $\approx$35 Oe for $t_{Co}$ = 3 \AA\, which set the maximum driving field at which the DW velocity could be measured.

The DW velocity was measured as a function of field, current, and temperature using a high-bandwidth scanning magneto-optical Kerr effect (MOKE) polarimeter.  The measurements tracked DW propagation along a 10-$\mu$m strip segment at timescales spanning up to 8 decades, following the procedure described in~\cite{emoriMOKE}.  For each measurement sequence, a reversed domain was initialized by the Oersted field from a current pulse through the transverse Cu nucleation line (Fig. \ref{fig1}(a)).  Then, an out-of-plane driving field $H$ expanded the reversed domain and drove a DW away from nucleation line (left to right in Fig. \ref{fig1}(a)).  In some measurements (Section \ref{sec5}), DW motion was assisted by an in-plane DC current $J_e$ injected through the Co/Pt strip.  The substrate temperature $T$ was controlled to an accuracy of $\pm$0.2 K with a thermoelectric module. 

For each driving parameter set ($H, J_e, T$), the MOKE transient signal was averaged over at least 100 cycles to account for the stochasticity of thermally activated DW motion.  These averaged MOKE transients (Figs. \ref{fig1}(b) and (c)) represent probability distributions for magnetization switching due to DW motion.  The insets of Figs. \ref{fig1}(b) and (c) show the average DW arrival time, taken as the time $t_{0.5}$ corresponding to 50\% switching probability, plotted against probed position along the strip.  The linear increase in $t_{0.5}$ versus position implies a uniform average DW velocity governed by a fine-scale disorder potential rather than discrete pinning sites.  The distinct profiles of the MOKE transients between Figs. \ref{fig1}(b) and (c) indicate a transition from stochastic to deterministic DW propagation as the driving field is increased beyond the strength of the pinning potential~\cite{emoriMOKE}. 

\section{\label{sec3}Field-Driven Domain Wall Motion}
\begin{figure}
\centering
\includegraphics[width=1\columnwidth]{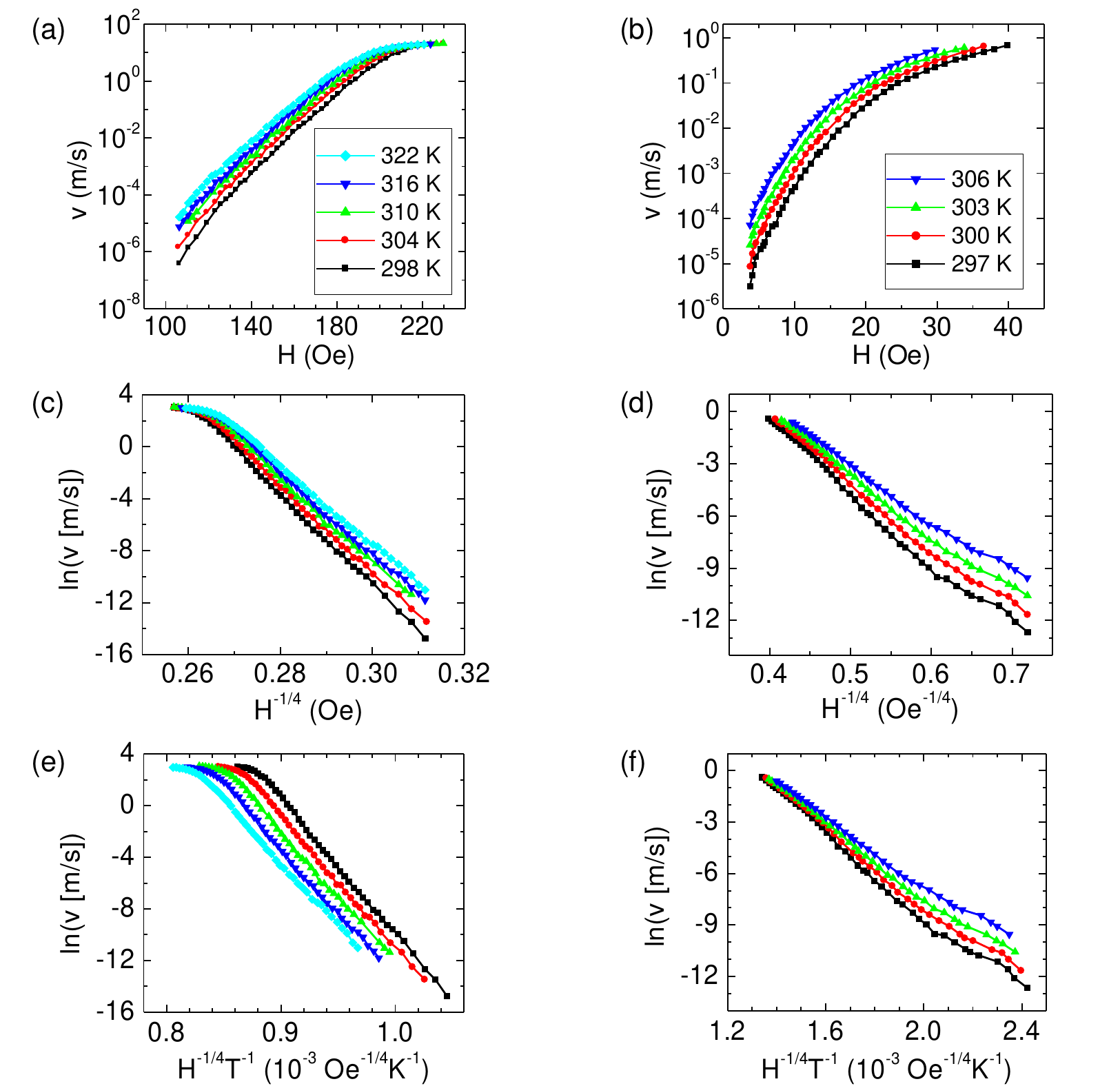}
\caption{\label{fig2}(a, b) DW velocity $v$ as a function of driving field $H$ at zero current at several different sample temperatures for the $t_{Co}$ = 7 \AA\ (a) and 3 \AA (b) samples. (c, d) ln($v$) plotted against $H^{-1/4}$ in accordance with the conventional creep scaling $\mu = 1/4$ for the $t_{Co}$ = 7 \AA\ (c) and 3 \AA (d) samples.  (e, f) ln($v$) plotted against $H^{-1/4}T^{-1}$ showing the failure of the conventional creep scaling for the $t_{Co}$ = 7 \AA\ (e) and 3 \AA (f) samples.}
\end{figure}

The DW velocity with zero driving current was measured over a range of driving field and several sample temperatures.  The 7-\AA\ sample exhibited substantial pinning strength so that a driving field of over 100 Oe was required to move DWs at $v > 10^{-6}$ m/s (Fig. \ref{fig2}(a)). By contrast, DW motion was detected at fields $< 10$ Oe in the 3-\AA\ strip (Fig. \ref{fig2}(b)).  Here, $H$ was adjusted, taking into account the small out-of-plane remanent field of the electromagnet core, to generate linear ln($v$) versus $H^{-1/4}$ relations in Fig. \ref{fig2}(b).  As shown in Fig. \ref{fig2}, the measured DW velocity spanned several decades.  The velocity at low driving fields increased by more than an order of magnitude with a small temperature difference of 24 K for $t_{Co}$ = 7 \AA\ and 9 K for $t_{Co} $= 3 \AA.  This large change in the DW velocity with both the driving field and sample temperature is a key characteristic of a thermally activated process. 
            
In Figs. \ref{fig2}(c) and (d), the DW velocity curves are plotted against $H^{-1/4}$, following the conventional creep scaling in which the activation energy in Eq. \ref{eq1} scales as $E_A \sim H^{-\mu}$  with the creep exponent $\mu = 1/4$~\cite{lemerle, cayssol, metaxas, kimNat, lavrijPin, mooreHighSpeed, alvarez, leePRL, cormierJPD, lavrijPt}.  Each mobility curve at a fixed temperature appears to be well described by a constant slope in ln($v$) versus $H^{-1/4}$ over several decades of velocity.  For the creep scaling of $\mu = 1/4$ to be valid, Eq. \ref{eq1} implies that the mobility curves at different temperatures should collapse on top each other if they are re-plotted as ln($v$) versus $H^{-1/4}T^{-1}$.  However, this is clearly not the case, as the mobility curves do not collapse for either $t_{Co} = 7$ \AA\ (Fig. \ref{fig2}(e)) or $t_{Co} = 3$ \AA\ (Fig. \ref{fig2}(f)).  The conventional creep scaling therefore does not apply in any range of the driving field examined in this study.

\begin{figure}
\centering
\includegraphics[width=1\columnwidth]{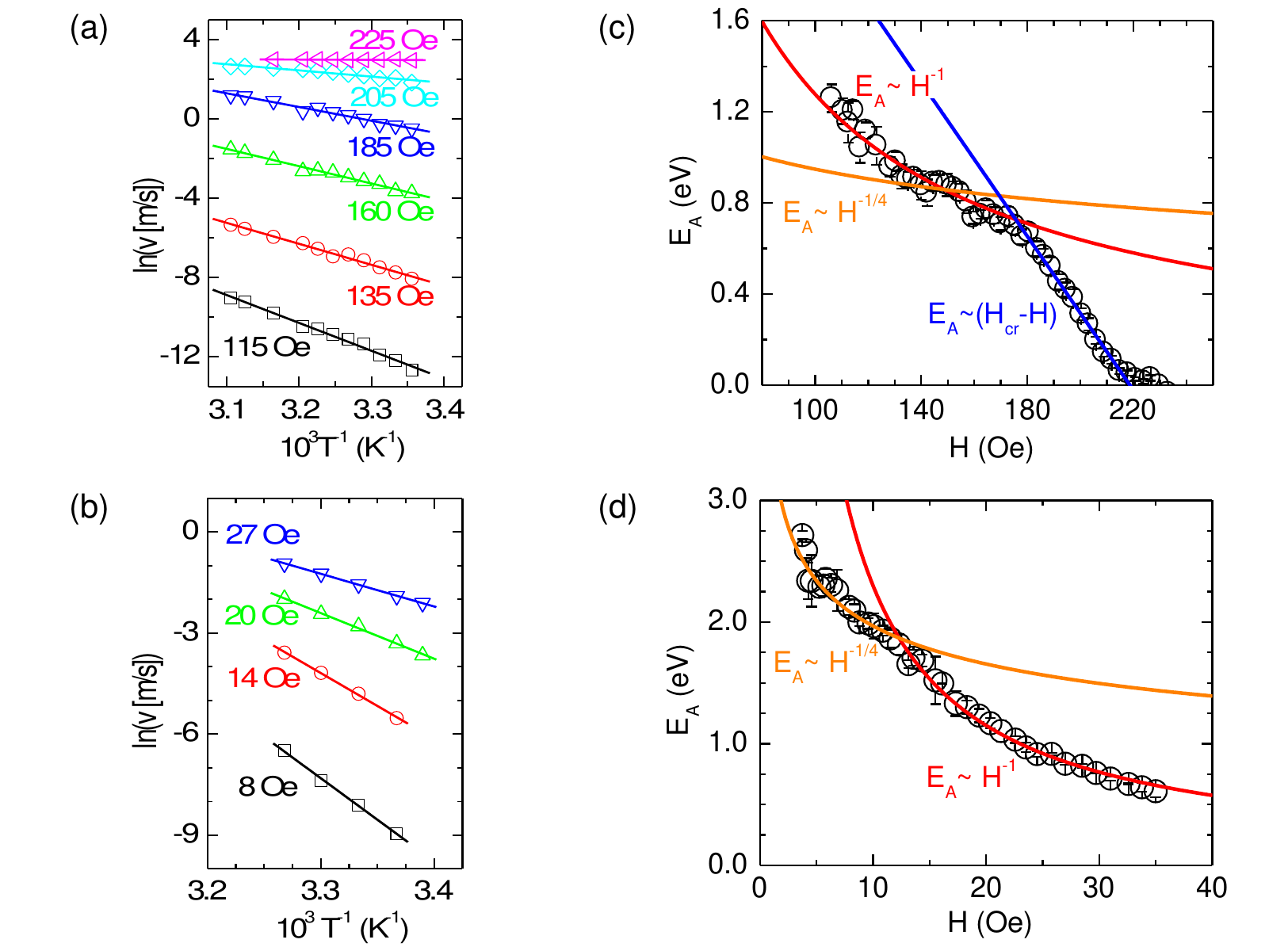}
\caption{\label{fig3}(a, b) Extraction of activation energy by fitting DW velocity against inverse temperature for the $t_{Co}$ = 7 \AA\ (a) and 3 \AA\ (b) samples. (c, d) Plot of activation energy versus driving field for the $t_{Co}$ = 7 \AA\ (c) and 3 \AA\ (d) samples.}    
\end{figure}

Instead of attempting to adjust the scaling exponents to better collapse the data, we directly extracted the activation energy from the slope of ln($v$) versus $T^{-1}$  (Figs. \ref{fig3}(a) and (b)).  The validity of the linear fit confirms our assumption that the activation energy is invariant with temperature in the narrow measured range of $\sim$10 K, although a Curie-temperature-dependent correction needs to be incorporated for generalized analyses as described in Section \ref{sec4}.  The values of the activation energy plotted in Figs. \ref{fig3}(c) and (d) are of similar magnitude with those reported previously for DW depinning~\cite{kimEa} and continuous DW motion~\cite{emoriJPCM, emoriGdOx}, and decrease monotonically with increasing field as expected.  
  
However, for $t_{Co}$ = 7 \AA, $E_A$ does not follow a simple creep power law scaling with $\mu = 1/4$, despite the linear isothermal ln($v$) versus $H^{-1/4}$ curves in Fig. \ref{fig2}(c).  In the lower field range ($H < 180$ Oe), $E_A$ scales instead with an exponent $\mu = 1$, which corresponds to the ``random-field" universality class similar to the experimental and theoretical findings in a GaMnAs ferromagnetic semiconductor~\cite{yamanouchi}, but different from $\mu = 1/4$ (``random-bond") widely used in previous studies of Co/Pt~\cite{lemerle, cayssol, metaxas, kimNat, lavrijPin, mooreHighSpeed, alvarez, leePRL, cormierJPD, lavrijPt}.  Due to its larger thickness, the 7-\AA\ Co/Pt multilayer possibly does not behave as a two-dimensional DW medium, a prerequisite to the creep dynamics with $\mu = 1/4$~\cite{lemerle}.  
The scaling of $E_A$ with $H$ in the 3-\AA\ strip cannot be described by a single power-law exponent, indicating that a single model of creep scaling cannot accurately capture the DW dynamics. These results suggest that in real samples, the defect potential is likely more complex than can be accounted for by a simple scaling model.  
Nonetheless, the directly extracted activation energy curves in Figs. \ref{fig3}(c) and (d) can be taken as the fingerprint of DW interactions with defects, which allows the thermally activated dynamics to be described without initially assuming any particular scaling model. We will show that this empirical approach, developed further in Section \ref{sec4}, allows for a direct quantitative assessment of the influence of current on DW motion as shown in Section \ref{sec5}.

At a sufficiently large driving force, the general scaling of creep breaks down and the activation energy vanishes.  Although such breakdown of creep could not be observed for the 3-\AA\ sample due to spontaneous nucleation of magnetic domains at $H \approx 35$ Oe, the DW dynamics of the 7-\AA\ sample indeed deviates significantly from creep above $H \approx 180$ Oe.  Here, the activation energy follows the simple linear relationship of depinning~\cite{ferre, emoriJPCM}, described by $E_A = 2V_A M_S (H_{cr}-H_{eff})$, where $V_A$  is the activation volume,  $M_S$ is the saturation magnetization, and $H_{cr}$ is the effective critical depinning field.  Taking $M_S$ = 1700 emu/cm$^3$~\cite{emoriCoPt, linCoPt, knepper},  $H_{cr}$ = 220 Oe,  $2V_A M_S$ = 0.017 eV/Oe (slope from Fig. \ref{fig3}(b)), and the effective magnetic thickness $t_{eff} = 3t_{Co} =$ 2.1 nm, we obtain a characteristic activation length $L_A = \sqrt{V_A/t_{eff}} \approx$ 62 nm, in line with $\approx$50 nm estimated for a similar Co/Pt multilayer strip with $t_{Co} = 6$ \AA~\cite{emoriJPCM}.  The activation energy vanishes at $H_{cr}$ = 220 Oe, above which DW motion is governed by viscous flow rather than thermal activation.  This is consistent with the findings in~\cite{emoriMOKE, metaxas, mironDW} in which the DW dynamics became viscous and deterministic above a critical field.  The transition from creep to depinning to viscous flow is seen directly in the activation energy as a function of driving field (Fig. \ref{fig3}(c)), whereas it is not evident in the DW velocity at a fixed temperature (Fig. \ref{fig2}(a)).   
      
\section{\label{sec4}Generalized Empirical Arrhenius-like Equation}
\begin{figure}
\centering
\includegraphics[width=1\columnwidth]{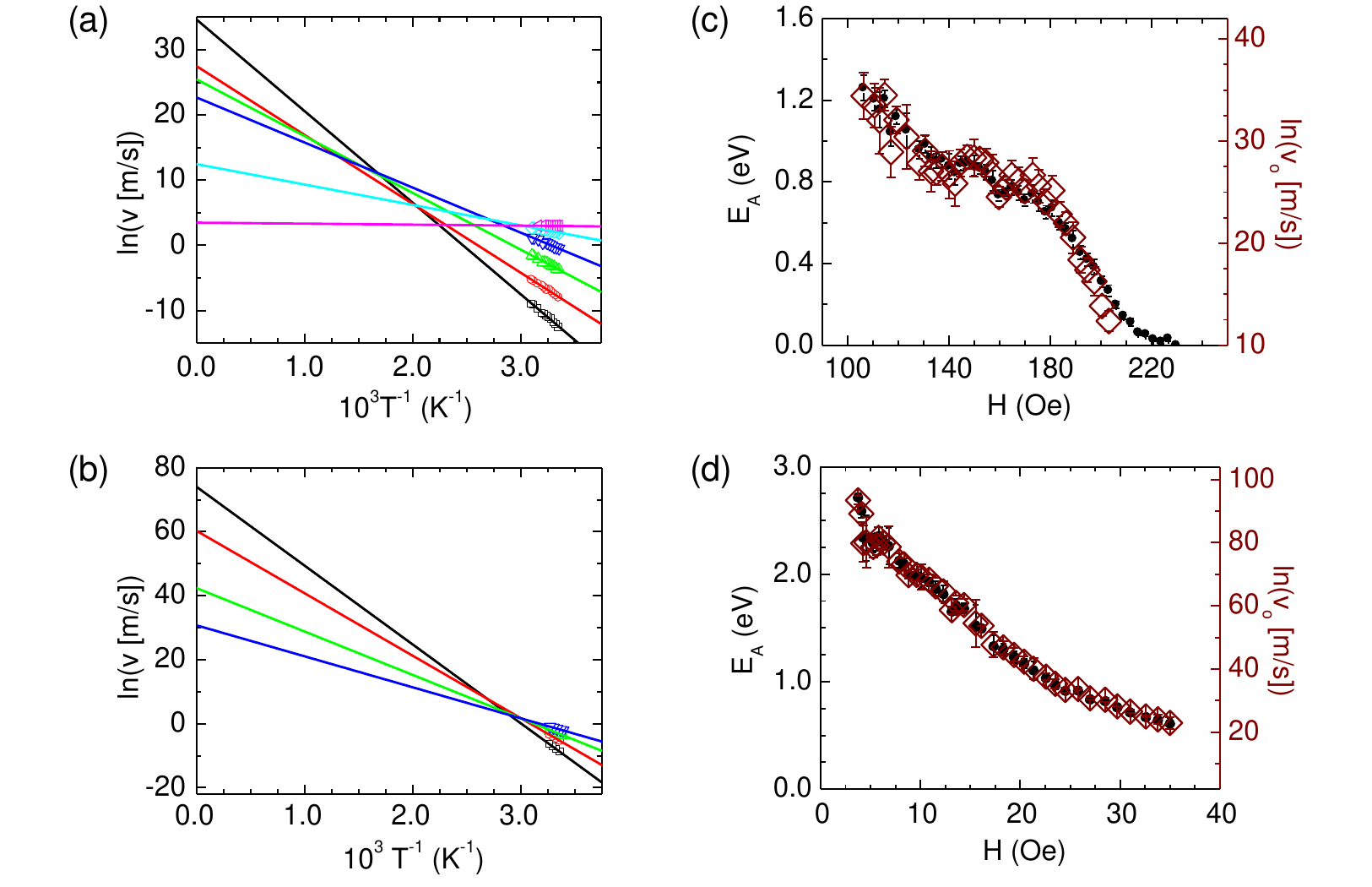}
\caption{\label{fig4}(a, b) Arrhenius plots displaying the same data as Fig. \ref{fig3}, showing the extrapolated fit lines do not converge at $T^{-1} = 0$ for both the (a) 7-\AA\ and (b) 3-\AA\ samples.  (c, d)  Dependence of $ln(v_o)$ with $H$, which tracks the dependence of $E_A$ with $H$, for both (c) 7-\AA\ and (d) 3-\AA\ samples.}    
\end{figure}

The pre-exponential factor $v_o$ in the Arrhenius relationship (Eq. \ref{eq1}) is extracted from the intercept of the Arrhenius plot at $T^{-1}$ = 0.  In the analysis described previously, the pre-exponential was assumed to be a constant with respect to driving field.  However, the extrapolated fit lines of ln($v$) versus $T^{-1}$ at different driving fields shown in Figs. \ref{fig4}(a) and (b) diverge significantly at $T^{-1}$ = 0, indicating clearly that the intercept ln($v_o$) is not constant.  Interestingly, as shown in Figs. \ref{fig4}(c) and (d), ln($v_o$) and the activation energy plotted as a function of the driving field track each other almost perfectly when they are scaled linearly with respect to each other.  The relationship between ln($v_o$) and the activation energy can be written as
\begin{equation}
\ln(v_o) = a+bE_A\label{eq2},
\end{equation}
where $a$ and $b$ are linear scaling constants.  Substituting Eq. \ref{eq2} into Eq. \ref{eq1}, we derive a generalized Arrhenius-like equation
\begin{equation}
v = A\exp\left(\frac{-E_A}{k_B T}\left[1-\frac{T}{T_{cr}}\right]\right)\label{eq3},
\end{equation}
where $T_{cr} = 1/bk_B$ and $A = \exp(a)$ is the corrected pre-exponential constant independent of driving field and temperature.  

Here, $E_A$ is the effective activation energy within a narrow range of measured temperature such that the Arrhenius plot produces an approximately constant slope, as shown in Fig. \ref{fig3}.  In reality, the activation energy scales with the saturation magnetization $M_S$ and must vanish at the Curie temperature $T_C$.  The corrected activation energy therefore should be ${E_A}^* = E_A (1-T/T_C)$, which is satisfied by Eq. \ref{eq3} if $T_C = T_{cr}$.   

\begin{figure}
\centering
\includegraphics[width=0.5\columnwidth]{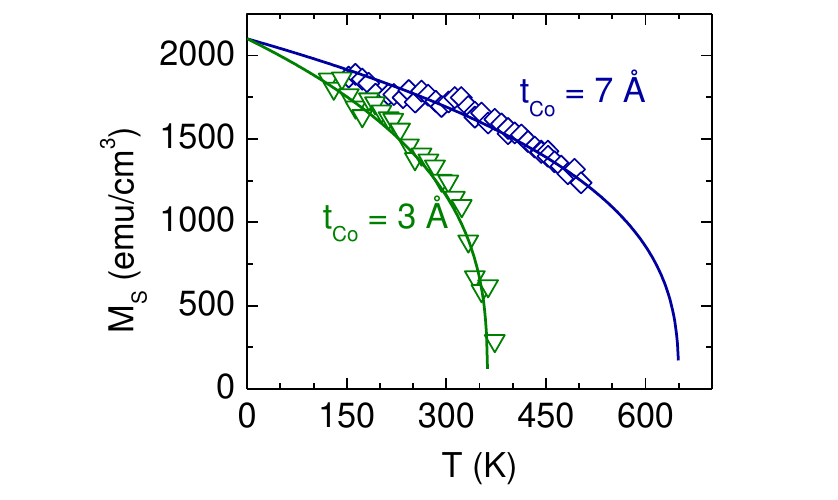}
\caption{\label{fig5}Saturation magnetization as a function of temperature.  The solid curves show fits to estimate the Curie temperature for each sample.}    
\end{figure}

We estimated $T_C$ by measuring the temperature dependence of $M_S$ with vibrating sample magnetometry on continuous Co/Pt multilayer films.  $M_S$ in Fig. \ref{fig5}, normalized with respect to the net nominal Co volume, is larger than that of bulk Co, which is typically $\sim$1400 emu/cm$^3$ at room temperature.  The larger $M_S$ in the Co/Pt multilayers is likely due to additional magnetization induced in Pt interfaced with Co, which has also been reported in \cite{emoriCoPt, linCoPt, knepper}. 
By fitting the data in Fig.~\ref{fig5} with a scaling of the form $(1-T/T_C)^\gamma$, where $\gamma$ is an empirical fitting exponent, $T_C$ of the 7-\AA\ and 3-\AA\ samples were estimated to be 620$\pm$70 and 360$\pm$20 K, respectively.  By comparison, the scaling of ln($v_o$) and $E_A$ (Fig. \ref{fig4} and Eq. \ref{eq2}) yielded $T_{cr}$ = 580 and 340 K for the 7-\AA\ and 3-\AA\ samples, respectively.  Since these two independently measured quantities are in reasonable agreement with each other, we conclude that $T_{cr} \approx T_C$ and the corrected activation energy for DW motion is ${E_A}^* \approx E_A (1-T/T_C)$.

\section{\label{sec5}Current-Assisted Domain Wall Motion}
Using the generalized Arrhenius-like equation derived in the previous section, we investigated the roles of current on DW motion.  To distinguish thermal and spin-torque effects, we first quantified Joule heating in the Co/Pt strips.  By measuring the electrical resistance with respect to current density $J_e$ and substrate temperature $T_{sub}$ set by the thermoelectric module, the sample temperature was found to increase quadratically with $J_e$ from Joule heating as  $\Delta T = h J_e^2$, where $h$ = 2.5 and 2.2 K/[10$^{11}$ A/m$^2$]$^2$ for the 7-\AA\ and 3-\AA\ samples, respectively.  In extracting the activation energy, the actual temperature of the sample was used so that $T = T_{sub}+\Delta T$.  

Current-driven DW motion was assisted with background driving fields, whose values were chosen to investigate each distinct dynamic regime found from field-driven DW measurements (Fig. \ref{fig3}).  The driving electron current density $J_e$ was estimated by assuming a uniform current distribution across the total conductive (Co and Pt) cross-sectional area, and we defined $J_e > 0$ when electron flow was in the same direction as field-driven DW motion (left to right in Fig. \ref{fig1}(a)).  The maximum $|J_e|$ was limited to $< 10^{11}$ A/m$^2$ to minimize electromigration in the multilayers.       
                              
We now incorporate the effects of injected current in the Arrhenius-like equation (Eq. \ref{eq3}): 
\begin{equation}
v = A\exp\left(\frac{-E_A(H_{eff})}{k_B [T_{sub}+hJ_e^2]}\left[1-\frac{T_{sub}+hJ_e^2}{T_{cr}}\right]\right)\label{eq4}.
\end{equation}
Here, the activation energy is expressed a function of the effective driving field $H_{eff} = H + \epsilon J$, where $\epsilon$ is the spin-torque efficiency, and the sample temperature includes the Joule heating effect.  This is a complete equation for thermally activated DW motion where the DW velocity is a function of field, current, and substrate temperature.  Table \ref{tab1} lists the empirical parameters from the measurements of field-driven DW motion (Section \ref{sec4}).  These parameters were then substituted into Eq. \ref{eq4}, and the spin-torque efficiency $\epsilon$ was extracted by fitting Eq. \ref{eq4} to the DW velocity data.  

\begin{figure}
\centering
\includegraphics[width=1\columnwidth]{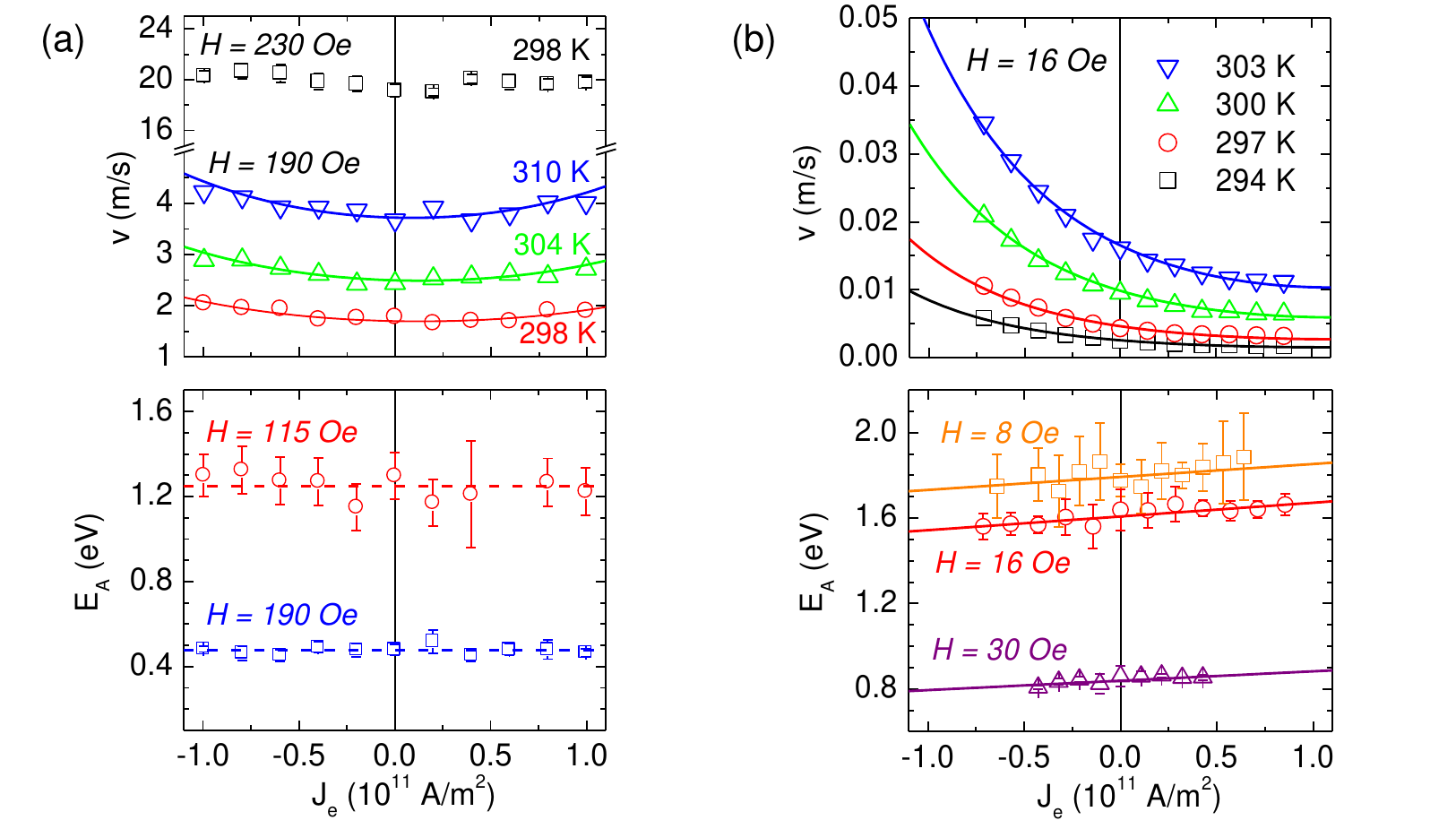}
\caption{\label{fig6}DW velocity and activation energy versus current density for $t_{Co}$ = 7 \AA\ (a) and $t_{Co}$ = 3 \AA\ (b).  With $J_e > 0$, the electron flow is in the same direction as field-driven DW motion. Solid curves in the velocity plots are fits using the canonical equation (Eq. \ref{eq4}) with the spin-torque efficiency $\epsilon$ adjusted.}    
\end{figure}

For the 7-\AA\ sample, the DW velocity increased by only ~10\% in both polarities in the thermally activated regime (e.g. $H$ = 190 Oe in Fig. \ref{fig6}(a)) and exhibited no systematic change in the viscous flow regime ($H$ = 230 Oe).   Fitting the velocity data with Eq. \ref{eq4} reveals that $\epsilon$ is at most -0.2 Oe/10$^{11}$ A/m$^2$, with the negative sign indicating that DW motion was facilitated slightly against the direction of electron flow.  This vanishingly small current-induced effect was verified with the activation energy for DW motion, which shows no systematic variation with current density (Fig. \ref{fig6}(a)).  Thus, in the 7-\AA\ sample, current increases the DW velocity through Joule heating but generates a negligible spin-torque effect.  Pinning alone cannot explain the very low spin-torque efficiency because no significant current-induced effect on DW motion was observed even when the pinning was nullified at $H$ = 230 Oe.  The lack of spin-torque effects in this Co/Pt multilayer instead arises from a vanishingly small effective spin-polarized current, as shown rigorously by Cormier \emph{et al}.~\cite{cormierPRB}
               
On the other hand, as shown in Fig. \ref{fig6}(b), a clear shift in the DW velocity with $J_e$ was observed in the 3-\AA\ sample.  This significant change is similar to the exponential relationship between the DW velocity and driving field (Fig. \ref{fig2}(b)), suggesting that current generates a spin torque on a DW that can be equated to an out-of-plane field.  Fitting of Eq. \ref{eq4} to the DW velocity data produced a spin-torque efficiency of $\epsilon$ = -2.6$\pm$-0.8 Oe/10$^{11}$ A/m$^2$.   Despite this clear current-dependence on the DW velocity, the variation of the activation energy with current density $\Delta E_A/\Delta J_e$ was at most 0.1 eV/10$^{11}$ A/m$^2$ (Fig. \ref{fig4}(b)), which is too small to elucidate any nonlinearity in the activation energy~\cite{kimEa} (see further discussion in Section \ref{sec7}) for the narrow range of current density in this present study.  The activation energy as a function of current density $J_e$ at a fixed driving field $H_o$ can be approximated to first order, assuming a small $\epsilon J_e$ so that $H_{eff} \approx H$:
\begin{eqnarray}
E_A(H_{eff})|_{H_o} &=& E_A(J_e)|_{H_o} 
\approx E_A|_{H_o}+\left.\frac{dE_A}{dH_{eff}}\right|_{H_o}\epsilon J_e \nonumber \\
&\approx& E_A|_{H_o}+\left.\frac{dE_A}{dH}\right|_{H_o} \epsilon J_e
\label{eq5}.
\end{eqnarray}	
Therefore, the spin torque efficiency can be estimated by
\begin{equation}
\epsilon \approx \left. \frac{\Delta E_A}{\Delta J_e} \right/ \left. \frac{\partial E_A}{\partial H} \right|_{H_o}
\label{eq6},
\end{equation}			
where ${\partial E_A}/{\partial H}|_{H_o}$ was calculated from the experimental data (Table I).  Using Eq. \ref{eq6}, we obtain $\epsilon$ = -1.2$\pm$0.6 Oe/10$^{11}$ A/m$^2$. The spin-torque efficiency in the Pt/(Co/Pt)$_3$ multilayer with $t_{Co}$ = 3 \AA\ is on the same order of magnitude as the typical efficiency in permalloy~\cite{vernier}, but about an order of magnitude smaller than the efficiencies reported by some prior studies on Co/Pt structures~\cite{ravelosona, boulle, alvarez, leePRL, lavrijPt}.   

\section{\label{sec6}Origin of Current-Induced Torques in Co/Pt}
Current-induced DW motion has usually been attributed to adiabatic and nonadiabatic spin-transfer torques (STTs)~\cite{ralph, beach, brataas, zhang, thiavilleSTT, tatara}.  Adiabatic STT drives DWs in thick ($\gg$ 1 nm) out-of-plane magnetized structures, e.g. Co/Ni mulitlayers~\cite{koyamaNatMat, ueda, tanigawa, koyamaHindep, kimTwoBarrier}, and its symmetry is distinct from an external magnetic field.  The mechanism of nonadiabatic STT is controversial, but its symmetry is known to be equivalent to an external field that drives a DW ~\cite{thiavilleSTT, zhang}.  The magnitudes of adiabatic and nonadiabatic STTs scale with the spin polarization of in-plane current, which can be large in thick Co/Ni~\cite{ueda, tanigawa}, but typically small in Co/Pt due to spin scattering by Pt~\cite{cormierPRB, houssam}.  If either of these conventional STTs were responsible for driving DWs in Co/Pt multilayers, the spin-torque efficiency would be larger in the 7-\AA\ sample, which is expected to carry a greater spin-polarized current with the thicker ferromagnetic Co layers.  Our experimental findings are contrary to this expectation: the Co/Pt multilayer with atomically thin (3 \AA) Co layers exhibits a finite spin-torque effect, whereas the multilayer with the larger Co thickness (7 \AA) does not.  The spin polarization and therefore conventional STTs are likely very small in these Co/Pt multilayers, as reported previously by a number of studies~\cite{mooreHighSpeed, mironSTMeter, suzuki, cormierPRB, cormierJPD, lavrijPt, emoriJPCM, emoriGdOx}.  The current-assisted DW motion in the ultrathin 3-\AA\ structure does not arise from nonadiabatic STT, but rather from another spin-torque mechanism equivalent to an effective out-of-plane magnetic field.  

We also note that in the 3-\AA\ structure, DW motion is assisted against the direction of electron flow ($J_e < 0$), opposite to the direction induced by conventional STTs.  A theoretically predicted negative spin polarization~\cite{sipr} or nonadiabatic STT coefficient~\cite{bohlens} may drive DWs against electron flow, but neither of these is applicable if the absolute magnitude of the spin polarization is vanishingly small.  The Oersted field is not responsible for this anomalous direction of motion, because we observed the same current-polarity dependence for both possible magnetization configurations across the DW (down-up and up-down).  DW charging by the extraordinary Hall effect~\cite{viret} is negligible in this metallic Co/Pt strip, and hydromagnetic DW drag~\cite{viret} should move DWs in the direction of the electron flow.  

Current-assisted DW motion opposing electron flow has also been reported in asymmetric Pt/Co/Pt trilayers with thicker Pt at the bottom layer~\cite{kimAPEX, leePRL, lavrijPt} and in Pt/Co(ferromagnet)/oxide(insulator) trilayers with Pt at the bottom~\cite{mooreHighSpeed, mironDW, emoriGdOx, emoriChiral, ryuChiral, emoriDMI, koyamaMgO, ryuParkin}.  A nonuniform current distribution due to the asymmetric layer structure may produce an internal electric potential, which then may generate an effective Rashba field in the ultrathin ferromagnet~\cite{mironRashba, wangManchon}.  Certain combinations of nonadiabatic STT and torques due to the Rashba field have been shown theoretically to move a DW against electron flow~\cite{kimRashba}.  However, the Rashba field scales with the spin polarization of current~\cite{mironRashba, wangManchon, kimRashba}, so it cannot explain the anomalous direction of motion under a small spin polarized current.  

The only known physical phenomenon that may explain the observed DW dynamics in the 3-\AA\ Co/Pt sample is the spin Hall effect (SHE)~\cite{liuPt, liuTa, haazen}, which generates an out-of-plane spin current from an in-plane charge current in a nonmagnetic heavy metal with strong spin-orbit coupling (e.g. Pt).  The torque from the SHE exerted on magnetic moments does not depend on the spin polarization of in-plane current through the ferromagnet.  Recent studies~\cite{khval, thiavilleDMI, haazen, emoriChiral, ryuChiral, emoriDMI} have shown that the SHE drives DWs in out-of-plane magnetized structures, if these DWs have an internal magnetization component longitudinal to the strip length direction, i.e. in the N\'eel configuration.  The effective field from the SHE acting on a N\'eel DW points out-of-plane~\cite{khval, emoriChiral, emoriDMI}, thereby producing an apparent equivalence between current- and field-driven DW motion.  The SHE thus explains the reports of robust current-induced DW motion resembling nonadiabatic STT in various Co/Pt-based structures, even though the spin polarization may be vanishingly small in these ultrathin Co layers.   Furthermore, depending on the chirality of N\'eel DWs and the sign of the spin Hall angle in the heavy metal, these DWs may move in either direction with respect to electron flow~\cite{emoriChiral, emoriDMI}.

The SHE exerts a torque on N\'eel DWs~\cite{khval} but not on Bloch DWs that are magnetostatically preferred in typical $\sim$500-nm wide strips~\cite{koyamaNatMat}.  Haazen \emph{et al}.~\cite{haazen} applied an in-plane longitudinal magnetic field to force DWs to a N\'eel configuration and attain SHE-assisted motion in asymmetric Pt/Co/Pt.  In our experiment, because no in-plane longitudinal field was applied, intrinsic properties of the Co/Pt multilayers must stabilize N\'eel DWs.  N\'eel DWs have been shown to be stable in Au/(Co/Au)$_2$ multilayers due to the closure of the stray field between the coupled ferromagnetic layers~\cite{bellec}.  Such unique magnetostatics within multilayers might have likewise stabilized N\'eel DWs in our Pt/(Co/Pt)$_3$.  Moreover, Bandiera \emph{et al}.~\cite{bandieraIEEE, bandieraAPL} show that the properties of the top and bottom interfaces in Pt/Co/Pt are significantly different, with the bottom Pt-Co interface exhibiting stronger interfacial magnetic anisotropy.  Our Co/Pt multilayers may possess similar structural asymmetry, leading to a moderate Dzyaloshinskii-Moriya interaction (DMI)~\cite{thiavilleDMI, fert, heide, chen, chenNat} that could stabilize N\'eel DWs with a fixed chirality~\cite{emoriChiral, ryuChiral, emoriDMI}.  A recent experiment indeed shows that such homochiral N\'eel DWs, stabilized by the interfacial DMI, are present even in nearly symmetric Pt/Co/Pt~\cite{jeDMI}.

We also note that the layer structures of Co/Pt were nominally symmetric (Section \ref{sec2}), so one might expect zero net spin current from the SHE.  However, the distribution of charge current might have been nonuniform due to interfacial scattering in the thin ($<$ 1.6 nm) Pt layers, thereby producing a finite SHE-induced spin current.  The ultrathin 3-\AA\ Co layers probably resulted in the combination of a N\'eel DWs and a small SHE torque, sufficient to generate a spin-torque efficiency of magnitude $\sim$1 Oe/10$^{11}$ A/m$^2$.   The spin-torque efficiency can be as high as $\sim$10 Oe/10$^{11}$ A/m$^2$ in Co/Pt structures with unequal top and bottom Pt layer thicknesses~\cite{alvarez, leePRL, lavrijPt} and approaches $\sim$100 Oe/10$^{11}$ A/m$^2$ in Pt/Co/insulator structures~\cite{mironDW, mironSTMeter, emoriGdOx, emoriChiral, emoriDMI, koyamaMgO}.  
These findings illustrate the possibility of engineering extremely efficient current-induced DW motion in heavy-metal/ultrathin-ferromagnet structures as an alternative to devices relying on conventional STTs.    

\section{\label{sec7}Nonlinear Current-Induced Effects}
Current may influence thermally activated DW motion in a number of different ways.  As described in Sections \ref{sec5} and \ref{sec6}, the SHE (or a current-induced effective out-of-plane field) modifies the activation energy unidirectionally similar to an applied field (see Eq. \ref{eq5}), whereas Joule heating enhances DW motion irrespective of current polarity through a temperature rise that enters directly into the Arrhenius relation (Eq. \ref{eq4}).   Theoretical studies~\cite{tatara, kimJV, ryu, duine} have also predicted that adiabatic STT decreases the activation energy for motion independent of current polarity, by distorting the DW configuration within the pinning potential.  This effect from adiabatic STT can be modeled as a quadratic reduction in the activation energy with respect to current~\cite{ryu}, which was reported in a recent experimental study~\cite{kimEa}.  We show that the generalized Arrhenius-like equation (Eq. \ref{eq4}), which incorporates the temperature dependence of the activation energy, provides an alternative explanation for this quadratic contribution through Joule heating.  Specifically, Joule heating can affect DW dynamics by reducing the saturation magnetization, which has been considered in a few studies of high-speed DW motion~\cite{ueda, curiale} but never for thermally activated motion.  

In ~\cite{kimEa}, the activation energy $E_A^{dep}$  was obtained from the DW depinning time $\tau$ at a single pinning site, such that $E_A^{dep}=k_B T \ln(f_o \tau)$, where $f_o = 10^9$ Hz is the attempt frequency.  After Joule heating $T = T_{sub} + hJ_e^2$ was included in the analysis, there still remained a clear quadratic change in the activation energy with current density $E_A \sim J_e^2$, which was attributed to adiabatic STT.  This quadratic contribution from current can originate from reduced saturation magnetization if the measured depinning energy barrier is of the form
\begin{equation}
E_A^{dep}=E_A \left(1-\frac{T}{T_{cr}}\right)\label{eq7},
\end{equation}
which takes into account the temperature dependence of the saturation magnetization (Section \ref{sec4}).  Here, assuming that current generates an effective field through the SHE (quantified by the spin-torque efficiency $\epsilon$), we approximate $E_A^{dep}$ at a fixed driving field $H_o$ by substituting Eq. \ref{eq5} into Eq. \ref{eq7}:
\begin{eqnarray}
E_A^{dep}(J_e)|_{H_o}\approx E_A|_{H_o} \left(1-\frac{T_{sub}}{T_{cr}}\right) 
+\left. \frac{dE_A}{dH} \right|_{H_o} \left(1-\frac{T_{sub}}{T_{cr}}\right) \epsilon J_e \nonumber \\
-\frac{E_A|_{H_o}h}{T_{cr}}J_e^2 - \left. \frac{dE_A}{dH} \right|_{H_o} \frac{\epsilon h}{T_{cr}}J_e^3
\label{eq8}.\
\end{eqnarray}
With $E_A|_{H_o} \sim$ 1 eV, $h \sim 2\times10^{-22}$ K m$^4$/A$^2$, and $T_{cr} \sim$ 500 K, the coefficient of the quadratic term in Eq. \ref{eq8} is $\sim 10^{-25}-10^{-24}$  eV m$^4$/A$^2$.  This estimated magnitude is on the same order as the quadratic coefficient extracted in~\cite{kimEa}, where the maximum Joule heating was 30 K at $|J_e| = 3.5\times10^{11}$ A/m$^2$.  Therefore, a quadratic dependence of the activation energy with current may be accounted for by Joule heating, instead of adiabatic STT.  The current-induced reduction of the saturation magnetization, and hence the activation energy, may not be trivial and should be considered in rigorous analyses of thermally activated DW motion. 

\section{\label{conc}Conclusion}
The analysis of the thermal activation energy barrier allows for directly extracting the scaling and effectiveness of field and current in moving a DW.  We have characterized the activation energy for DW motion driven by both field and current in Co/Pt multilayers with $t_{Co}$ = 7 \AA\ and 3 \AA.  The distinct dynamic regimes of creep, depinning, and viscous flow are resolved by the field dependence of the activation energy.  For a consistent Arrhenius-like description, the activation energy must incorporate a correction factor dependent on the Curie temperature of the sample.  One must be careful with quantifying the effects of driving current, which affects thermally activated DW motion not only directly through temperature rise and spin torques, but also indirectly through a decrease in the saturation magnetization.  

By incorporating the effects of Joule heating, we have derived a generalized equation for thermally activated DW motion as a function of field, current, and temperature.  The use of this generalized equation reveals that current generates a negligible torque on DWs in the Co/Pt multilayer with thick (7 \AA) Co layers.  By contrast, the multilayer with thin (3 \AA) Co layers exhibits a spin-torque efficiency large enough to assist DW motion against electron flow.  These results suggest that current-induced DW motion in these Co/Pt multilayers does not require any spin polarization of in-plane current through the ferromagnet, but instead may depend on the Co-Pt interfaces and the magnitude of net spin current generated from the spin Hall effect in Pt.  

\section*{Acknowledgements}
S.E. acknowledges financial support by the NSF Graduate Research Fellowship.  C.K.U. was supported by the MIT Undergraduate Research Opportunities Program.  Instruments in the Center of Materials Science and Engineering, the Scanning-Electron-Beam Lithography facility at the Research Laboratory of Electronics, and the Nanostructures Laboratory at MIT were used to fabricate the samples for this study.




\begin{thebibliography}{00}


\bibitem{ferre} J. Ferr\'{e}, \textit{Spin Dynamics in Confined Magnetic Structures I}, (Heidelberg: Springer, 2002). 
\bibitem{lemerle} S. Lemerle, J. Ferr\'e, C. Chappert, V. Mathet, T. Giamarchi, and P. Le Doussal, Phys. Rev. Lett. {80} (1998) 849.
\bibitem{cayssol} F. Cayssol, D. Ravelosona, C. Chappert, J. Ferr\'e, and J. P. Jamet, Phys. Rev. Lett. {92} (2004) 107202.
\bibitem{metaxas} P. J. Metaxas, J. P. Jamet, A. Mougin, M. Cormier, J. Ferr\'e, V. Baltz, B. Rodmacq, B. Dieny, and R. L. Stamps, Phys. Rev. Lett. {99} (2007) 217208.
\bibitem{kimNat} K.-J. Kim, J.-C. Lee, S.-M. Ahn, K.-S. Lee, C.-W. Lee, Y. J. Cho, S. Seo, K.-H. Shin, S.-B. Choe, and H.-W. Lee, Nature {458} (2009) 740 .
\bibitem{lavrijPin} R. Lavrijsen et al., Appl. Phys. Lett. {98} (2011) 132502. 
\bibitem{emoriMOKE} S. Emori, D. C. Bono, and G. S. D. Beach, J. Appl. Phys. {111} (2012) 07D304.
\bibitem{jeDMI}S.-G. Je, D.-H. Kim, S.-C. Yoo, B.-C. Min, K.-J. Lee, and S.-B. Choe, Phys. Rev. B. {88} (2013) 214401.
\bibitem{wuth}C. Wuth, P. Lendecke, and G. Meier, J. Phys: Cond. Matter {24} (2012) 024207.
\bibitem{allwood}D. A. Allwood, G. Xiong, C. C. Faulkner, D. Atkinson, D. Petit, and R. P. Cowburn, Science {309} (2005) 1688.
\bibitem{kiermaier}J. Kiermaier, S. Breitkreutz, I. Eichwald, M. Engelstädter, X. Ju, G. Csaba, D. Schmitt-Landsiedel, and M. Becherer, J. Appl. Phys. {113} (2013) 17B902.
\bibitem{parkin}S. S. P. Parkin, M. Hayashi, and L. Thomas, Science {320} (2008) 190.
\bibitem{currivan}J. Currivan, Y. Jang, M.D. Mascaro, M.A. Baldo, and C.A. Ross, Magn. Lett. IEEE {3} (2012) 3000104.
\bibitem{koyamaNatMat} T. Koyama, D. Chiba, K. Ueda, K. Kondou, H. Tanigawa, S. Fukami, T. Suzuki, N. Ohshima, N. Ishiwata, Y. Nakatani, K. Kobayashi, and T. Ono, Nat. Mater. {10} (2011) 194.
\bibitem{ueda} K. Ueda, T. Koyama, R. Hiramatsu, D. Chiba, S. Fukami, H. Tanigawa, T. Suzuki, N. Ohshima, N. Ishiwata, Y. Nakatani, K. Kobayashi, and T. Ono, Appl. Phys. Lett. {100} (2012) 202407.
\bibitem{tanigawa}H. Tanigawa, T. Suzuki, S. Fukami, K. Suemitsu, N. Ohshima, and E. Kariyada, Appl. Phys. Lett. {102} (2013) 152410.
\bibitem{ngo} D.-T. Ngo, K. Ikeda, and H. Awano, Appl. Phys. Express {4} (2011) 093002.  
\bibitem{koyamaHindep} T. Koyama, D. Chiba, K. Ueda, H. Tanigawa, S. Fukami, T. Suzuki, N. Ohshima, N. Ishiwata, Y. Nakatani, and T. Ono, Appl. Phys. Lett. {98} (2011) 192509.
\bibitem{mooreHighSpeed} T. A. Moore,  I. M. Miron, G. Gaudin, G. Serret, S. Auffret, B. Rodmacq, A. Schuhl, S. Pizzini, J. Vogel and M. Bonfim, Appl. Phys. Lett. {93} (2008) 262504. 
\bibitem{mironDW} I. M. Miron, I. M. Miron, G. Gaudin, G. Serret, S. Auffret, B. Rodmacq, A. Schuhl, S. Pizzini, J. Vogel, and M. Bonfim, Nat. Mat. {10} (2011) 419.
\bibitem{ryuParkin}K.-S. Ryu, L. Thomas, S.-H. Yang, and S. S. P. Parkin, Appl. Phys. Express {5} (2012) 093006.
\bibitem{koyamaMgO}T. Koyama, H. Hata, K.-J. Kim, T. Moriyama, H. Tanigawa, T. Suzuki, Y. Nakatani, D. Chiba, and T. Ono, Appl. Phys. Express {6} (2013) 033001.
\bibitem{ravelosona} D. Ravelosona, D. Lacour, J. A. Katine, B. D. Terris, and C. Chappert, Phys. Rev. Lett. {95} (2005) 117203.  
\bibitem{boulle} O. Boulle, J. Kimling, P. Warnicke, M. Kl\"aui, U. R\"udiger, G. Malinowski, H. J. M. Swagten, B. Koopmans, C. Ulysse, and G. Faini, Phys. Rev. Lett. {101} (2008) 216601.   
\bibitem{burrowes} C. Burrowes, A.P. Mihai, D. Ravelosona, J.-V. Kim, C. Chappert, L. Vila, A. Marty, Y. Samson, F. Garcia-Sanchez, L.D. Buda-Prejbeanu, I. Tudosa, E.E. Fullerton, and J.-P. Attane, Nat Phys {6} (2010) 17.
\bibitem{alvarez} L. San Emeterio Alvarez, , K.-Y. Wang, S. Lepadatu, S. Landi, S. J. Bending, and C. H. Marrows, Phys. Rev. Lett. {104} (2010) 137205.
\bibitem{heinen} J. Heinen, O. Boulle, K. Rousseau, G. Malinowski, M. Kl\"aui, H. J. M. Swagten, B. Koopmans, C. Ulysse, and G. Faini, Appl. Phys. Lett. {96} (2010) 202510. 
\bibitem{kimAPEX} K.-J. Kim, J.-C. Lee, S.-J. Yun, G.-H. Gim, K.-S. Lee, S.-B. Choe, and K.-H. Shin, Appl. Phys. Express {3} (2010) 083001.
\bibitem{leePRL} J.-C. Lee, K.-J. Kim, J. Ryu, K.-W. Moon, S.-J. Yun, G.-H. Gim, K.-S. Lee, K.-H. Shin, H.-W. Lee, and S.-B. Choe, Phys. Rev. Lett. {107} (2011) 067201.
\bibitem{mironSTMeter} I. M. Miron, P.-J. Zermatten, G. Gaudin, S. Auffret, B. Rodmacq, and A. Schuhl, Phys. Rev. Lett. {102} (2009) 137202. 
\bibitem{suzuki} T. Suzuki, S. Fukami, K. Nagahara, N. Ohshima, and N. Ishiwata, IEEE Trans. Magn. {44} (2008) 25. 
\bibitem{cormierPRB} M. Cormier, A. Mougin, J. Ferr\'e, A. Thiaville, N. Charpentier, F. Pi\'echon, R. Weil, V. Baltz, and B. Rodmacq, Phys. Rev. B {81} (2010) 024407.
\bibitem{cormierJPD} M. Cormier, A. Mougin, J. Ferr\'e, J.-P. Jamet, R. Weil, J. Fassbender, V. Baltz, and B. Rodmacq, J. Phys. D: Appl. Phys. {44} (2011) 215002.
\bibitem{lavrijPt}R. Lavrijsen, P. P. J. Haazen, E. Mur\`e, J.H. Franken, J. T. Kohlhepp, H. J. M. Swagten, and B. Koopmans, Appl. Phys. Lett. {100} (2012) 262408.
\bibitem{kimEa} K.-J. Kim, , J. Ryu, G.-H. Gim, J.-C. Lee, K.-H. Shin, H.-W. Lee, and S.-B. Choe, Phys. Rev. Lett. {107} (2011) 217205.
\bibitem{kimDHtemp}D.-H. Kim, K.-W. Moon, S.-C. Yoo, B.-C. Min, K.-H. Shin, and S.-B. Choe, IEEE Trans. Magn.  {49} (2013) 3207. 
\bibitem{kimTwoBarrier}K.-J. Kim, R. Hiramatsu, T. Koyama, K. Ueda, Y. Yoshimura, D. Chiba, K. Kobayashi, Y. Nakatani, S. Fukami, M. Yamanouchi, H. Ohno, H. Kohno, G. Tatara, and T. Ono, Nat. Commun. {4} (2013) 2011.
\bibitem{emoriJPCM} S. Emori and G. S. D. Beach, J. Phys: Cond. Matter {24} (2012) 024214.
\bibitem{emoriGdOx} S. Emori, D.C. Bono, and G.S.D. Beach, Appl. Phys. Lett. {101} (2012) 042405.
\bibitem{yamanouchi} M. Yamanouchi, J. Ieda, F. Matsukura, S. E. Barnes, S. Maekawa, and H. Ohno, Science {317} (2007) 1726.  
\bibitem{curiale}J. Curiale, A. Lemaitre, C. Ulysse, G. Faini, and V. Jeudy, Phys. Rev. Lett. {108} (2012) 076604.
\bibitem{haazen}P. P. J. Haazen, E. Mur\`e, J. H. Franken, R. Lavrijsen, H. J. M. Swagten, and B. Koopmans, Nat. Mater. {12} (2013) 299.
\bibitem{emoriChiral} S. Emori, U. Bauer, S.-M. Ahn, E. Martinez, and G. S. D. Beach, Nat. Mater. {12} (2013) 611.
\bibitem{ryuChiral}K.-S. Ryu, L. Thomas, S.-H. Yang, and S. Parkin, Nat. Nanotechnol. {8} (2013) 527.
\bibitem{emoriDMI}S. Emori, E. Martinez, U. Bauer, S.-M. Ahn, P. Agrawal, D. C. Bono, and G. S. D. Beach, arXiv:1308.1432 (2013).
\bibitem{martinezStoch}E. Martinez, J. Phys. Condens. Matter 24 (2012) 024206.
\bibitem{ralph} D. C. Ralph and M. D. Stiles, J. Magn. Magn. Mater. {320} (2008) 1190.
\bibitem{beach} G. S. D. Beach, M. Tsoi, and J. L. Erskine, J. Magn. Magn. Mater. {320} (2013) 1272.
\bibitem{brataas}A. Brataas, A. D. Kent, and H. Ohno, Nat. Mater. {11} (2012) 372.
\bibitem{zhang} S. Zhang and Z. Li, Phys. Rev. Lett. {93} (2004) 127204.
\bibitem{thiavilleSTT}A. Thiaville, Y. Nakatani, J. Miltat, and Y. Suzuki, Europhys. Lett. {69} (2005) 990.
\bibitem{tatara}G. Tatara, T. Takayama, H. Kohno, J. Shibata, Y. Nakatani, and H. Fukuyama, J. Phys. Soc. Jpn. {75} (2006) 064708.
\bibitem{khval}A. V. Khvalkovskiy, V. Cros, D. Apalkov, V. Nikitin, M. Krounbi, K. A. Zvezdin, A. Anane, J. Grollier, and A. Fert, Phys. Rev. B {87} (2013) 020402.
\bibitem{kimJV} J.-V. Kim and C. Burrowes, Phys. Rev. B. {80} (2009) 214424. 
\bibitem{ryu} J. Ryu, S.-B. Choe, and H.-W. Lee, Phys. Rev. B. {84} (2011) 075469.  
\bibitem{liuPt}L. Liu, O. J. Lee, T. J. Gudmundsen, D. C. Ralph, and R. A. Buhrman, Phys. Rev. Lett. {109} (2012) 096602.
\bibitem{liuTa}L. Liu, C.-F. Pai, Y. Li, H. W. Tseng, D. C. Ralph, and R. A. Buhrman, Science {336} (2012) 555.
\bibitem{thiavilleDMI}A. Thiaville, S. Rohart, É. Jué, V. Cros, and A. Fert, Europhys. Lett. {100} (2012) 57002.
\bibitem{emoriCoPt} S. Emori and G. S. D. Beach, J. Appl. Phys. {110} (2011) 033919. 
\bibitem{linCoPt}C.-J. Lin, G. L. Gorman, C. H. Lee, R. F. C. Farrow, E. E. Marinero, H. V. Do, H. Notarys, and C. J. Chien, J. Magn. Magn. Mater. {93} (1991) 194.
\bibitem{knepper}J. W. Knepper and F. Y. Yang, Phys. Rev. B. {71} (2005) 224403.
\bibitem{vernier} N. Vernier, D. A. Allwood, D. Atkinson, M. D. Cooke, and R. P. Cowburn, Europhys. Lett. {65} (2004) 7.
\bibitem{houssam}D. Houssameddine, U. Ebels, B. Delaët, B. Rodmacq, I. Firastrau, F. Ponthenier, M. Brunet, C. Thirion, J.-P. Michel, L. Prejbeanu-Buda, M.-C. Cyrille, O. Redon, and B. Dieny, Nat. Mater. {6} (2007) 447.
\bibitem{sipr} O. \v{S}ipr, J. Minar, S. Mankovsky, and H. Ebert, Phys. Rev. B {78} (2008) 144403.
\bibitem{bohlens} S. Bohlens and D. Pfannkuche, Phys. Rev. Lett. {105} (2010) 177201.
\bibitem{viret} M. Viret, A. Vanhaverbeke, F. Ott, and J.-F. Jacquinot, Phys. Rev. B {72} (2005) 140403(R).
\bibitem{mironRashba}  I. M. Miron, G. Gaudin, S. Auffret, B. Rodmacq, A. Schuhl, S. Pizzini, J. Vogel, and P. Gambardella, Nat. Mat. {9} (2010) 230.
\bibitem{wangManchon}X. Wang and A. Manchon, Phys. Rev. Lett. {108} (2012) 117201.
\bibitem{kimRashba}K.-W. Kim, S.-M. Seo, J. Ryu, K.-J. Lee, and H.-W. Lee, Phys. Rev. B {85} (2012) 180404.
\bibitem{bellec}A. Bellec, S. Rohart, M. Labrune, J. Miltat, and A. Thiaville, Europhys. Lett. {91} (2010) 17009.
\bibitem{bandieraIEEE}S. Bandiera, R. R. Sousa, B. B. Rodmacq, and B. Dieny, Magn. Lett., IEEE {2} (2011) 3000504.
\bibitem{bandieraAPL}S. Bandiera, R. C. Sousa, B. Rodmacq, and B. Dieny, Appl. Phys. Lett. {100} (2012) 142410.
\bibitem{fert}A. Fert, V. Cros, and J. Sampaio, Nat. Nanotech. {8} (2013) 152.
\bibitem{heide}M. Heide, G. Bihlmayer, and S. Bl\"ugel, Phys. Rev. {78} (2008) 140403.
\bibitem{chen}G. Chen, J. Zhu, A. Quesada, J. Li, A.T. N'Diaye, Y. Huo, T.P. Ma, Y. Chen, H.Y. Kwon, C. Won, Z.Q. Qiu, A.K. Schmid, and Y.Z. Wu, Phys. Rev. Lett. 110 (2013) 177204.
\bibitem{chenNat} G. Chen, T.P. Ma, A.T. N'Diaye, H.Y. Kwon, C. Won,	Y.Z. Wu, and A.K. Schmid, Nat. Comm. {4} (2013) 2671. 
\bibitem{duine} R. A. Duine and C. M. Smith, Phys. Rev. B. {77} (2008) 094434. 

\end{thebibliography}


\newpage

\clearpage

\clearpage

\clearpage

\clearpage

\clearpage

\clearpage
\begin{center}
\begin{table*}[b]
\caption{\label{tab1}Parameters in the generalized Arrhenius-like equation (Eq. \ref{eq4}) to quantify effects of current on DW motion} 
\centering 
\begin{tabular}{c ccccc} 
\hline\hline 
$t_{Co}$&\multicolumn{2}{c}{7 \AA}&\multicolumn{2}{c}{3 \AA} \\ [0.5ex] 
\hline 
{} & $H < $ 180 Oe & $H > $ 180 Oe& $H < $ 13 Oe& $H > $ 13 Oe\\
$E_A$ (eV) & $CH_{eff}^{-1}$ & $C(H_{cr}-H_{eff})$ &  $CH_{eff}^{-1/4}$ &  $CH_{eff}^{-1}$\\ 
$C$ & 130 eV Oe & 0.017 eV/Oe & 3.5 eV Oe$^{-1/4}$ & 23 eV Oe\\
$H_{cr}$ (Oe) & --- & 220 & --- & --- \\
$T_{cr}$ (K) & \multicolumn{2}{c}{580} & \multicolumn{2}{c}{340}\\
$A $(m/s) & \multicolumn{2}{c}{$2.2\times10^4$} & \multicolumn{2}{c}{7.4}\\
$h$(K/[10$^{11}$A/m$^2$]$^2)$& \multicolumn{2}{c}{2.5} & \multicolumn{2}{c}{2.2}\\[1ex] 
\hline 
\end{tabular}
\label{tab:hresult}
\end{table*}
\end{center}

\end{document}